\documentclass[seceq]{ptptex}
\usepackage{amsmath,amssymb}
\usepackage{graphicx}

\newcommand{\beq}{\begin{equation}}
\newcommand{\beqa}{\begin{eqnarray}}
\newcommand{\eeq}{\end{equation}}
\newcommand{\eeqa}{\end{eqnarray}}
\newcommand{\non}{\nonumber}
\newcommand{\lb}{\label}
\newcommand{\fr}[1]{(\ref{#1})}
\newcommand{\cc}{\mbox{c.c.}}

\newcommand{\ve}{{\varepsilon}}
\newcommand{\e}{\mbox{e}}
\newcommand{\tA}{\widetilde A}
\newcommand{\tB}{\widetilde B}
\newcommand{\tC}{\widetilde C}
\newcommand{\tv}{\widetilde v}
\newcommand{\ta}{\widetilde a}
\newcommand{\tb}{\widetilde b}
\newcommand{\cA}{{\cal A}}
\newcommand{\tcA}{\widetilde{\cal A}}
\notypesetlogo                       

\title{
Renormalization Reductions for Systems with Delay
}

\author{
Shin-itiro \textsc{Goto} 
}

\inst{
NTT Communication Science Laboratories, NTT Corporation\\
Seika-cho, Kyoto 619-0237, Japan
}

\recdate{someday; Submitted to Prog. Theor. Phys.}

\abst{
The renormalization method which is a type of perturbation method  
is extended to a tool to study weakly nonlinear time-delay systems.  
For systems with order-one delay, 
we show that the renormalization method leads to reduced systems without delay.
For systems with order-one and large-delay,
we propose an extended renormalization method which 
leads to reduced systems with delay.  
In some examples, the validities of our perturbative results 
are confirmed analytically and numerically. 
We also compare our reduced equations with reduced ones 
obtained by another perturbation method.
}

\begin{document}

\maketitle
\section{Introduction}
\lb{sec:intro}
Many nonlinear dynamical systems in various scientific disciplines
are influenced by the finite propagation time of signals in feedback loops.
A typical physical system is provided by a laser system where the output light
is reflected and fed back to the cavity\cite{LK80,PM03,USNY04}.
Time delays also occur in other situations. 
For example,
in a traffic flow model including a driver's reaction time\cite{New61}, 
in biology due to physiological control mechanisms\cite{MG77}, or 
in economy where the finite velocity of information processing has to be  
taken into account\cite{Mac89}. Furthermore, realistic models in population
dynamics or in ecology include the duration for the replacement of the 
resources\cite{Gop92}. 
In some situations, such as lasers and electro-mechanical systems\cite{JM99}, 
systems with large-delay appear.
For this reason, 
we need to develop a mathematical tool to study them, especially for 
weakly nonlinear systems as a first step.
The main difficulty peculiar to systems with delay is its dimensionality.
Due to a delayed arraignment in a given system, 
$x(t-r)=\exp(-r\partial/\partial t)x(t)$, the dimension of the phase 
space is high.

Suppose we add a perturbation term to a given system, 
the system is not guaranteed to be structurally stable. So 
the perturbation result is, if computed naively, plagued with singularities
such as secular terms. 
It has been recognized that these singularities in the result of the 
naive perturbation method can be 
renormalized away by the modification (renormalization) 
of parameters associated 
with the unperturbed system\cite{CGO}.
The modified parameters are governed by the renormalization equations that 
turn out to be slow-motion equations (reduced equations).
It is important that the prescription of the method does not depend on
the details of the system under study. 
To obtain a more 
useful and sophisticated tool to study weakly nonlinear systems,
reformulated versions\cite{Oon00,Kun,EFK00,MO01,NO01,IN06,GMN99} 
of the original method\cite{CGO} have been proposed.
It is noted that there are a variety of 
applications of renormalization methods 
to physical systems, such as plasma physics\cite{Tze04}, 
general relativity\cite{Nam} and 
quantum optics\cite{Fra97}, 
in addition to studies in standard nonlinear dynamical systems.
Although the reformulated version of the renormalization method 
that we employ here 
is easily applied to non-chaotic 
systems\cite{GMN99,RG-DE} and chaotic maps\cite{chaos-maps},
we do not know whether or not 
the renormalization method can be applied to systems with delay.

The purpose of this paper is to show that the reformulated 
renormalization method can be 
applied to weakly nonlinear  systems with delay. 
For systems with order-one delay, the method leads to reduced systems 
without delay.
For systems with large-delay, the method should be extended and the 
application of the extended method leads to reduced systems with delay. 
Our extended method can also be applied to systems with order-one delay,
the resultant reduced equations are different from those obtained by the
use of the conventional renormalization method.
As mentioned, a time-delay term of a given 
system makes the dimension of the phase space high. 
Even in such a case, our method can lead to reduced equations.  

The organization of this paper is as follows. 
In the next section(\S\ref{sec:conventional}), 
We show that our conventional renormalization method 
can lead to reduced equations for systems with order-one delay.  
The validities of our analyzes are shown.
In \S\ref{sec:extended},
we propose an extended version of our reformulated renormalization method so 
that we deal with a system with large-delay. The definition of 
large-delay is to be given in the beginning of the section.
We show that the extended method can also be applied to systems with order-one
delay. The validity of the extended method is also discussed.
Finally, in \S\ref{sec:conclusion}, we discuss the features of our methods 
and conclude our study.

\section{Conventional Renormalization Method}
\lb{sec:conventional}
In this section, using the conventional renormalization method, 
we first analyze a linear system that has an oscillatory solution, 
and show that our perturbative analysis 
is in agreement with the exact solution analytically.
By the conventional method we mean the method proposed in Ref.\citen{GMN99}.  
Next, we study some classes of weakly nonlinear systems using our 
renormalization method. Our classes include a nonlinear oscillator, 
a laser model, systems with many degrees of freedom, 
and spatially extended systems. 
In some examples, we show that our analyzes are valid by comparing with 
the numerical simulation or the previous studies.
\subsection{ Linear system}
\lb{subsec:linear}
To show that our renormalization method\cite{GMN99} 
can be applied to systems with delay, 
we consider the following system as an example,
\beq
\frac{d^2 x(t)}{dt^2}+\omega^2 x(t) +\ve x(t-r)=0, 
\lb{eqn:linear-DDE}
\eeq
where $\omega(\in\mathbb{R})$ is a parameter, $r(\in\mathbb{R})$ represents 
the time-delay, and $\ve(\in\mathbb{R})$ is a small parameter ($|\ve|\ll 1$).
In this system, there is an oscillatory solution that is
analytically expressible without any approximation. 
The exact solution is written as
\beq
x(t)=A\exp(it\sqrt{\omega^2-\ve})+\cc,
\lb{eqn:linear-exact}
\eeq
under the condition
\beq
r=\frac{\pi}{\sqrt{\omega^2 -\ve}}.
\lb{eqn:linear-cond}
\eeq
Here $\cc$ represents for the complex conjugate terms 
of the preceding expression, 
$A(\in\mathbb{C})$ is the integration constant.

Let us derive a perturbation solution of \fr{eqn:linear-DDE} using our 
renormalization method. In this perturbative analysis, we do not use the 
exact solution \fr{eqn:linear-exact}. 
As well as in the case of a differential equation without delay,
we first find the naive perturbation solution, 
$x(t)=x^{(0)}(t)+\ve x^{(1)}(t)+\ve^2 x^{(2)}(t)+{\cal O}(\ve^3)$.
This naive perturbation 
solution is obtained by solving the following equations,
\beqa
&&L x^{(0)}(t)=0,\quad L x^{(j)}(t)=-x^{(j-1)}(t-r),(j=1,2,...)\non\\
&&L x(t):=\bigg(\frac{d^2}{dt^2}+\omega^2\bigg)x (t).\non
\eeqa
The solutions are obtained as following
\beqa
x^{(0)}(t)&=&A\e^{i\omega t}+\cc,\non\\
x^{(1)}(t)&=&\frac{iA}{2\omega}t\e^{i\omega (t-r)}+\cc,\non\\
x^{(2)}(t)&=&\frac{-A}{8\omega^2}
\bigg(
t^2 -2rt+\frac{i}{\omega}t
\bigg)
\e^{i\omega (t-2r)}+\cc,\non
\eeqa
where $A(\in\mathbb{C})$ is the integration constant of the solution of the 
unperturbed system, $x^{(0)}(t)$. 
Note that the solutions $x^{(j)}(t), (j\geq 1)$ 
contain the terms 
const.$\exp(i\omega t)$ and const.$\exp(-i\omega t)$. 
We assume that these terms are 
included in $A\exp(i\omega t)$ and its complex conjugate term in $x^{(0)}(t)$. 
Apparently, the validity of the naive perturbation solution is invalid  
in the regime $t>{\cal O}(1/\ve)$, 
due to the secular terms ($\propto \ve t, \propto \ve^2 t^2 $ etc.).

The renormalization method removes the secular behavior in a systematic way.
We define the  renormalized variable $\tA(t)$ up to ${\cal O}(\ve^2)$,
\beq
\tA(t):=A+\ve  \frac{iA}{2\omega}t \e^{-i\omega r}
+\ve^2\frac{-A}{8\omega^2}\bigg(
t^2 -2rt+\frac{i}{\omega}t\bigg)
\e^{-2i\omega r}.
\lb{eqn:linear-RGT}
\eeq
Note that this definition is a form of a near-identity transformation at the 
constant $A$\cite{MO01}, and  that the naive perturbation solution is 
expressed in terms of the renormalized variable,
\beq
x(t)=\tA(t)\exp(i\omega t)+\cc +{\cal O}(\ve^3).
\lb{eqn:linear-RGAx}
\eeq
We construct the equation which $\tA(t)$ should satisfy perturbatively.
Such an equation is our renormalization equation.
From Eq. \fr{eqn:linear-RGT},  
we obtain the following two relations,
\beqa
\tA(t+\sigma)-\tA(t)&=&\ve\frac{iA}{2\omega}\sigma\e^{-i\omega r}\non\\
&&+\ve^2\frac{-A}{8\omega^2}
\bigg(2t\sigma+\sigma^2-2r\sigma+\frac{i}{\omega}\sigma\bigg)\e^{-2i\omega r}
+{\cal O}(\ve^3).
\lb{eqn:linear-diffA}
\eeqa
and 
\beq
A=\tA(t)-\ve\frac{i\tA(t)}{2\omega}t\e^{-i\omega r}+{\cal O}(\ve^2),
\lb{eqn:linear-IRGT}
\eeq
where $\sigma(\in\mathbb{R})$ is a parameter.
Substituting Eq.\fr{eqn:linear-IRGT} into Eq.\fr{eqn:linear-diffA}, we have
the approximate closed relation 
\beqa
&&\frac{\tA(t+\sigma)-\tA(t)}{\sigma}=
\ve\frac{i\tA(t)}{2\omega}\e^{-i\omega r}
+\ve^2\tA(t)
\bigg(
\frac{r}{4\omega^2}-\frac{i}{8\omega^3}
\bigg)\e^{-2i\omega r}
+{\cal O}(\sigma,\ve^3).\non
\eeqa
The renormalization equation is obtained in the limit $\sigma\to 0$ as
\beq
\frac{d\tA(t)}{dt}=\ve\frac{i\tA(t)}{2\omega}\e^{-i\omega r}
+\ve^2 \tA(t)
\bigg(
\frac{r}{4\omega^2}-\frac{i}{8\omega^3}
\bigg)\e^{-2i\omega r}.
\lb{eqn:linear-RGE}
\eeq
The solution of Eq.\fr{eqn:linear-RGE} is given by
\beqa
\tA(t)&=&\tA(0)\e^{\phi(t)},
\lb{eqn:linear-RGS}\\
\phi(t)&:=&t\Bigg\{
\ve\frac{i}{2\omega}\e^{-i\omega r}
+
\ve^2\bigg(\frac{r}{4\omega^2}-\frac{i}{8\omega^3}
\bigg)\e^{-2i\omega r}
\Bigg\}.
\non
\eeqa

To compare the solution of our renormalization method with the exact solution 
\fr{eqn:linear-exact}, we restrict ourselves to the approximate solution 
imposed on the condition \fr{eqn:linear-cond}. 
Using Eq. \fr{eqn:linear-cond}, we can rewrite Eq. \fr{eqn:linear-RGS} as  
\beq
\tA(t)=\tA(0)\exp
\bigg(
\ve\frac{-it}{2\omega}+\ve^2\frac{-it}{8\omega^3}+{\cal O}(\ve^3)
\bigg).
\lb{eqn:linear-sol-tA}
\eeq
In terms of $x(t)$,  we obtain the approximate solution   
using Eqs. \fr{eqn:linear-RGAx} and \fr{eqn:linear-sol-tA},
\beqa
x(t)&=&\tA(0)\exp
\Bigg\{
i t\bigg(\omega - \frac{\ve}{2\omega}-\frac{\ve^2}{8\omega^3}+{\cal O}(\ve^3)
\bigg)
\Bigg\}
+\cc +{\cal O}(\ve^3).
\lb{eqn:linear-RGcomp}
\eeqa
In fact, due to the relation 
$$
\sqrt{\omega^2 - \ve}=
\omega - \frac{\ve}{2\omega} - \frac{\ve^2}{8\omega^3}+{\cal O}(\ve^3),
$$
Eq. \fr{eqn:linear-RGcomp} is the same as Eq. \fr{eqn:linear-exact} 
up to ${\cal O}(\ve^2)$.

\subsection{Nonlinear single oscillator}
\lb{subsec:nonlinear}
Let us consider a nonlinear equation including a time-delay term, 
which is to show that 
our renormalization method can be applied to such a system.
The system which we study here is 
\beq
\frac{dx(t)}{dt}+\alpha x(t)+\beta x(t-r)=\ve(\gamma_1 x(t)-\gamma_3 x^3(t)),
\lb{eqn:nonlinear}
\eeq
where $\alpha,\beta,\gamma_1(\in\mathbb{R})$ and $\gamma_3(\in\mathbb{R})$ 
are parameters. The value of $r(\in\mathbb{R})$ represents the time-delay, and
$\ve(\in\mathbb{R})$ is the small parameter.
It is noted here 
that the unperturbed system has an analytically expressible oscillatory 
solution when the following condition is satisfied,
\beq
r=\frac{\arccos(-\alpha/\beta)}{\sqrt{\beta^2-\alpha^2}}, 
\quad(\beta^2>\alpha^2).
\lb{eqn:nonlinear-cond}
\eeq
We restrict ourselves to the case that this condition is satisfied.
The oscillatory solution of the unperturbed system is given by
$$
x^{(0)}(t)=A\e^{i\omega t}+\cc,
$$
where $\omega:=\sqrt{\beta^2-\alpha^2}$, 
$A(\in\mathbb{C})$ is the integration constant, and the relation 
$i\omega+\alpha=-\beta\exp(-i\omega r)$ is satisfied. 
Although we can analytically 
express the exact solution of the unperturbed system, 
it is difficult to express the exact solution analytically 
in the case $\ve\neq 0$. 
To investigate the effect of the perturbation term,    
we construct 
the renormalization equation for  Eq.\fr{eqn:nonlinear}.
The procedure to obtain the renormalization equation is the same as that 
described in \S\ref{subsec:linear}.

The naive perturbation solution,  
$x(t)=x^{(0)}(t)+\ve x^{(1)}(t) +{\cal O}(\ve^2)$,  is obtained 
by solving the following equations,
\beqa
&&
L_rx^{(0)}(t)=0,
\quad L_rx^{(1)}(t)=\gamma_1 x^{(0)}(t)-\gamma_3 x^{(0)3}(t),\non\\
&&
L_r x(t):=\frac{dx(t)}{dt}+\alpha x(t) +\beta x(t-r).\non
\eeqa
The solutions are given by
\beqa 
x^{(0)}(t)&=&A\e^{i\omega t}+\cc,\non\\
x^{(1)}(t)&=&t\frac{\gamma_1 A- 3\gamma_3|A|^2A}{1+r(\alpha+i\omega)}
\e^{i\omega t}+\cc.\non
\eeqa
The naive perturbation solution includes the secular term. 
To remove the secular term, we define the renormalized variable $\tA(t)$.
$$
\tA(t):=A+\ve t\frac{\gamma_1 A- 3\gamma_3|A|^2A}{1+r(\alpha+i\omega)}.
$$
The renormalization equation is obtained as
\beq
\frac{d\tA(t)}{dt}=
\ve\frac{\gamma_1 \tA(t)- 3\gamma_3|\tA(t)|^2\tA(t)}{1+r(\alpha+i\omega)}.
\non
\eeq
This system can be split into two parts: dynamics described by its 
amplitude and phase, 
\beqa 
\frac{dR^2}{dt}&=&-3\gamma_3 Q R^2\bigg(R^2-\frac{\gamma_1}{3\gamma_3}\bigg),
\lb{eqn:nonlinear-RG-amp}\\
\frac{d\phi}{dt}&=&\frac{3\ve\gamma_3\omega}{(1+r\alpha)^2+(r\omega)^2}
\bigg(R^2-\frac{\gamma_1}{3\gamma_3}\bigg).
\lb{eqn:nonlinear-RG-phase}
\eeqa
Here, $\tA(t):=R(t)\e^{i\phi(t)}$ and 
$Q:=2\ve(1+r\alpha)/\{(1+r\alpha)^2+(r\omega)^2\}$.
It turns out that, 
from Eqs. \fr{eqn:nonlinear-RG-amp} and \fr{eqn:nonlinear-RG-phase}, 
$R=R_*, (R_*:=\sqrt{\gamma_1/(3\gamma_3)})$ is a stable fixed point 
when $\gamma_1/\gamma_3>0$ and $\gamma_3Q>0$. 
It is thus expected that the limit-cycle oscillation appears when these
conditions are satisfied. 
The amplitude of this limit-cycle oscillation is given by $2R_*$ due to the  
relation, $x(t)\approx 2R(t)\cos(\omega t+\phi(t))$.
Fig.\ref{fig:nonlinear} shows that our analysis is valid. 
 
\begin{figure}[h]
\unitlength 1mm
\begin{picture}(120,40)
\put(50,5){\includegraphics[width=5.2cm]{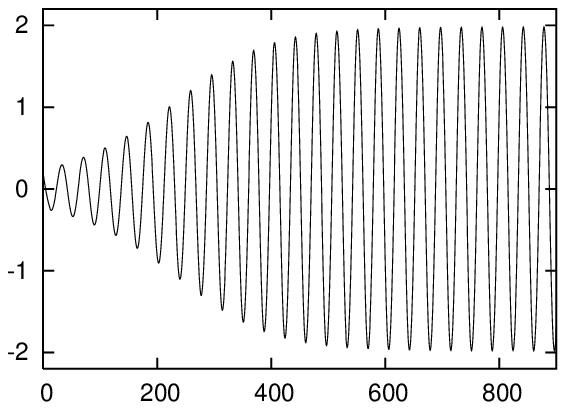}}
\put(75,0){$t$}
\put(40,25){$x(t)$}
\end{picture}
\caption{
The time sequence of the delay system \fr{eqn:nonlinear}. The values of the 
parameters are $\alpha=0.1,\beta=0.2,\gamma_1=3,\gamma_3=1$ and $\ve=0.01$.
The value of the time-delay $r$ is set to that of \fr{eqn:nonlinear-cond}.
The initial condition is $x(t)=0.2$ for $-r\leq t\leq 0$. 
The amplitude of the limit-cycle is  $2$, as predicted for this case 
[ $2R_*=2\sqrt{\gamma_1/(3\gamma_3)}=2$, see text ]. 
The numerical simulation was done 
using a fixed-step fourth-order Runge-Kutta method 
with linear interpolation 
for the required two midpoint evaluations of the delayed variable. 
}
\lb{fig:nonlinear}
\end{figure}

\subsection{Lang-Kobayashi phase equation}
In this subsection, we show that our analysis can lead to a set of 
reduced equations 
which describe dynamics of semiconductor lasers with feedback. 
In Ref.\citen{AKGE96}, they have analyzed the following delay equation
\beq
\Phi^{'''}+\omega\xi\Phi^{''}+\Phi^{'}-\Delta+\omega\Lambda_1\cos[\Phi(S-\Theta)-\Phi(S)] =0,
\lb{eqn:LK}
\eeq  
where $\xi,\Delta,\Lambda_1(\in\mathbb{R})$ are parameters, 
$\omega(\in\mathbb{R})$ is the small parameter, 
and the prime means differentiation with respect to $S$.
This equation is obtained 
from the Lang-Kobayashi equations in the following conditions\cite{AKGE96}:
the small ratio of the phonon and carrier lifetimes and the relatively 
large value of the linewidth enhancement factor.
We derive a reduced equation from Eq.\fr{eqn:LK} 
using our method and compare
our result with that obtained by using the multiple-scale method 
in Ref.\citen{AKGE96}. 

First, the naive perturbation solution, 
$\Phi(S)=\Phi^{(0)}(S)+\omega\Phi^{(1)}(S)+{\cal O}(\omega^2)$, is obtained by 
solving the following equations
\beqa 
L\Phi^{(0)}(S)&=&\Delta,
\lb{eqn:LK-0th}\\
L\Phi^{(1)}(S)&=&
-\xi\Phi^{(0)}(S)-\Lambda_1\cos[\phi^{(0)}(S-\Theta)-\Phi^{(0)}(S)] ,
\lb{eqn:LK-1st}\\
L\Phi(S)&:=&\bigg(\frac{d^3}{dS^3}+\frac{d}{dS}\bigg)\Phi(S).\non
\eeqa
The solution to the unperturbed system \fr{eqn:LK-0th}, $\Phi^{(0)}(S)$,  
is obtained as 
\beq
\Phi^{(0)}(S)=\frac{A}{2}\e^{i(S+v)}+\cc +S\Delta +B,
\lb{eqn:LK-0th-Sol}
\eeq
where $A,v(\in\mathbb{R})$, and $B(\in\mathbb{R})$ are the 
integration constants.
Substituting Eq. \fr{eqn:LK-0th-Sol} into  Eq. \fr{eqn:LK-1st}, we obtain the 
following equation
\beqa
L\Phi^{(1)}(S)&=&
-\xi A\cos(S+v)-\Lambda_1 J_0(D)\cos(\Theta\Delta)\non\\
&&-\Lambda_1\bigg\{
\cos(\Theta\Delta)\sum_{n=2,4,\cdots}J_n(D)\e^{in(-\Theta/2+S+v)}\non\\
&&\quad +\sin(\Theta\Delta)\sum_{n=1,3,\cdots}J_n(D)
\frac{\e^{in(-\Theta/2+S+v)}}{i}
\quad +\cc
\bigg\}.
\lb{eqn:LK-1st-sub}
\eeqa
Here $D$ is defined by $D:=2A\sin(\Theta/2)$,
$J_n$ denote the $n$-th order Bessel functions, and we have used the 
following relation in deriving Eq. \fr{eqn:LK-1st-sub},
$$
\e^{iz\sin\theta}=J_0(z)+2i\sum_{n=1,3,\cdots}J_n(z)\sin(n\theta)
+2\sum_{n=2,4,\cdots}J_n(z)\cos(n\theta).
$$
The solution of Eq.\fr{eqn:LK-1st} is given by
\beqa
\Phi^{(1)}(S)&=&-\frac{\xi A S}{2}\cos(S+v)
-\Lambda_1 J_0(D)S\cos(\Theta\Delta)\non\\
&&+S\Lambda_1 J_1(D)\sin(\Theta\Delta)\sin(-\Theta/2+S+v)\non\\
&&-\Lambda_1\bigg\{
\cos(\Theta\Delta)\sum_{n=2,4,\cdots}J_n(D)\frac{\e^{in(-\Theta/2+S+v)}}{in(1-n^2)}\non\\
&&\qquad 
-\sin(\Theta\Delta)\sum_{n=3,5,\cdots}J_n(D)\frac{\e^{in(-\Theta/2+S+v)}}{n(1-n^2)}+\cc
\bigg\}.
\lb{eqn:LK-1st-Sol}
\eeqa
This naive perturbation solution includes the secular terms 
($\propto \omega S$). 

Next, we remove these secular terms using our renormalization method.
We define the renormalized variables $\tC(S)(\in\mathbb{C})$ and
$\tB(S)(\in\mathbb{R})$ as follows
\beqa
\tC(S)&:=&A-\frac{\omega S}{2}
\bigg\{
\xi A+2i\Lambda_1 J_1(D(A))\sin(\Theta\Delta)\e^{-i\Theta/2} 
\bigg\},\non\\ 
\tB(S)&:=&B-\omega S\Lambda_1 J_0(D(A))\cos(\Theta\Delta).\non
\eeqa
The set of the renormalization equations 
up to ${\cal O}(\omega)$ is obtained as
\beqa
\frac{d\tC(S)}{dS}&=&-\frac{\omega}{2}\xi\tC(S)
-i\omega\Lambda_1 J_1(D(|\tC(S)|))
\sin(\Theta\Delta)\e^{-i\Theta/2},
\lb{eqn:LK-RG-C}\\
\frac{d\tB(S)}{dS}&=&-\omega\Lambda_1 J_0(D(|\tC(S)|))\cos(\Theta\Delta).
\lb{eqn:LK-RG-B}
\eeqa

Here, we compare the renormalization equations 
\fr{eqn:LK-RG-C} and \fr{eqn:LK-RG-B} with the 
reduced ones obtained by the multiple-scale method.   
Using the decomposition 
$\tC(S)=\tA(S)\e^{i\tv(S)},(\tA(S)\in\mathbb{R}, \tv(S)\in\mathbb{R})$,  
we obtain 
\beqa
\frac{d\tA(S)}{dS}&=&-\frac{\omega\xi}{2}\tA(S)
-\omega\Lambda_1\sin(\Theta\Delta)J_1(D(\tA(S)))\sin(\Theta/2),
\lb{eqn:LK-RG-A}\\
\frac{d\tv(S)}{dS}&=&-\frac{\omega\Lambda_1}{\tA(S)}\sin(\Theta\Delta)
J_1(D(\tA(S)))\cos(\Theta/2).
\lb{eqb:LK-RG-v}
\eeqa 
When we introduce the slow variable $\zeta:=\omega S$,
the renormalization equations \fr{eqn:LK-RG-B},\fr{eqn:LK-RG-A} and 
\fr{eqb:LK-RG-v} become the reduced equations 
derived in Ref.\citen{AKGE96}. This comparison 
shows that our analysis is consistent with one by a traditional 
perturbation method, and that our method can lead to 
the reduced equations from a physical system. 

\subsection{Weakly nonlinear lattice}
In this subsection, we show that our method leads to a discrete complex 
Ginzburg-Landau equation from a weakly nonlinear lattice with delay. 
In this lattice system, the finite propagation time of motion from the 
nearest oscillators is taken into account. 
Studying the derived reduced system, we predict the stability of 
a trivial solution, and this prediction is confirmed numerically.   

The weakly nonlinear oscillator which we study here is given by 
\beqa
\frac{dx_j(t)}{dt}&=&p_j(t),
\lb{eqn:WN_x}\\
\frac{dp_j(t)}{dt}&=&-\Omega^2 x_j(t)\non\\
&&+\ve\bigg\{ \nu\bigg(x_{j+1}(t-r)+x_{j-1}(t-r)-2x_j(t)\bigg)
 -\alpha x_j^3(t)
\bigg\},
\lb{eqn:WN_p}
\eeqa
where $\alpha,\nu(\in\mathbb{R})$ are parameters, $\ve(\in\mathbb{R})$ is the 
small parameter, and $r(\in\mathbb{R})$ represents the time-delay. 
The variables $x_j(t)(\in\mathbb{R})$ and $p_j(t)(\in\mathbb{R})$ 
denote the displacement and momentum of 
the single oscillator located at lattice site $j(\in\mathbb{Z})$ respectively.
It is noted that this given system becomes a Hamiltonian system when $r=0$.
Using the conventional renormalization method, we derive the reduced system 
here.

First, the naive perturbation solutions 
$x_j(t)=x_j^{(0)}(t)+\ve x_j^{(1)}(t)+{\cal O}(\ve^2)$ 
are obtained as
\beqa
x_j(t)&\approx& A_j\e^{i\Omega t}\non\\
&&+\ve\frac{t\e^{i\Omega t}}{2i\Omega}
\bigg[
\nu\{ \e^{-i\Omega r}(A_{j+1}+A_{j-1})-2A_j\}
-3\alpha|A_j|^2A_j\bigg]+\cc
\lb{eqn:WN_naive}
\eeqa
Here $A_j(\in\mathbb{C})$ are the integration constants, and 
the higher harmonic terms in $x_j^{(1)}(t)$ are omitted.

Second, from Eq.\fr{eqn:WN_naive}, 
the renormalized variables are defined as
\beqa
\tA_j(t)&:=&A_j+\ve 
\frac{t}{2i\Omega}
\bigg[
\nu\{ \e^{-i\Omega r}(A_{j+1}+A_{j-1})-2A_j\}
-3\alpha|A_j|^2A_j
\bigg].\non
\eeqa
From the definitions, we have the relation 
$x_j(t)\approx \tA_j(t)\e^{i\Omega t}+\cc$, and 
the renormalization equations 
\beqa
\frac{d\tA_j(t)}{dt}&=&\frac{\ve}{2i\Omega}
\bigg[
\nu\{ \e^{-i\Omega r}(\tA_{j+1}(t)+\tA_{j-1}(t))-2\tA_j(t)\}\non\\
&&\qquad
-3\alpha|\tA_j(t)|^2\tA_j(t)
\bigg].
\lb{eqn:WN_RGE}
\eeqa
The system \fr{eqn:WN_RGE} becomes the 
discrete nonlinear Shcr\"odinger equation, a Hamiltonian system,  
in the case $r=0$. 
When $r\neq 0$, Eq. \fr{eqn:WN_RGE} is the discrete complex Ginzburg-Landau 
equation.

In the rest of this subsection, 
we clarify a part of the phase space for the derived system \fr{eqn:WN_RGE}. 
Using the renormalization equations, 
we predict the behavior of motion in the original system 
and confirm it numerically.  
Here we restrict ourselves to the conditions $x_{j+N}(t)=x_j(t)$ with 
$N$ being the number of the oscillators, this conditions lead to 
$\tA_{j+N}(t)=\tA_j(t)$.
There is the trivial solution $\tA_j(t)=0$ in Eq. \fr{eqn:WN_RGE}. 
We show that the uniform solution, 
expressed as $\tA_j(t)\equiv\tA(t),$ (for any $j$),  
can be viewed as one of the local stable manifolds of 
the fixed point $\tA_j=0$ when a certain condition 
 is satisfied. To do this, we study the linear stability for $\tA_j=0$. 
Substituting $\tA_j(t)=\ta_j(t), (|\ta_j(t)|\ll 1)$ into Eq.\fr{eqn:WN_RGE} 
we have the linearized equation of motion in Fourier space
\beq
\frac{d\tb_k(t)}{dt}=\frac{-i\ve\nu}{\Omega}
\bigg\{
-1+\e^{-i\Omega r}\cos\bigg(\frac{2\pi k}{N}\bigg)
\bigg\}\tb_k(t),
\lb{eqn:WN_RGE_L}
\eeq
where 
$$
\tb_k(t)=\frac{1}{\sqrt{N}}\sum_{j=0}^{N-1}\e^{-i2\pi kj/N}\ta_j(t),\quad
\ta_j(t):=\frac{1}{\sqrt{N}}\sum_{k=0}^{N-1}\e^{i2\pi kj/N}\tb_k(t),
$$
with $(k=0,...,N-1)$. From Eq. \fr{eqn:WN_RGE_L}, we can predict which modes increase or decrease 
in time.  The absolute value of $b_k(t)$ decreases to zero as time evolves 
when 
\beq
\frac{\ve\nu}{\Omega}\sin(\Omega r)\cos\bigg(\frac{2\pi k}{N}\bigg)>0,
\lb{eqn:WN_cond_k}
\eeq
and this condition with $k=0$ gives that 
the uniform solution can be viewed as the local stable manifold of $\tA_j=0$. 
Fig. \ref{fig:lattice} shows that our analysis for the given system, via 
the renormalization equation, is valid.

\begin{figure}[h]
\unitlength 1mm
\begin{picture}(120,40)
\put(40,0){\includegraphics[width=7.0cm]{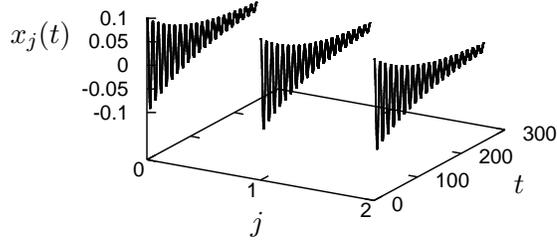}}
\put(65,5){$j$}
\put(33,30){$x_j(t)$}
\put(100,10){$t$}
\end{picture}
\caption{
The time sequence of the system described by Eqs.\fr{eqn:WN_x} and 
\fr{eqn:WN_p}. The values of the 
parameters are $\Omega=0.5,\alpha=1,\nu=1.01, r=1$ and $\ve=0.01$. 
The number of the oscillators $N$ is three, and $x_{j+N}(t)=x_j(t)$.
The initial conditions are $x_j(t)=0.1, p_j(t)=0,(j=0,1,2)$ 
for $-r\leq t\leq 0$, 
which correspond to the uniform solution.  
The amplitudes of $x_j(t),(j=0,1,2)$ decrease to zero as time evolves, 
which we can predict using the condition \fr{eqn:WN_cond_k} 
with $k=0$ [ See text ].
The numerical simulation method is given in the caption to 
Fig.\ref{fig:nonlinear}.  
}
\lb{fig:lattice}
\end{figure}

\subsection{Spatially extended system (1)}
To show that our renormalization method is also useful 
in the case that a spatially extended system modeled by 
including unperturbed terms with delayed arguments,
we consider the following system,
\beq
\frac{\partial u(t,x)}{\partial t}+\frac{\pi}{2r}u(t-r,x)
=\ve\bigg(\alpha u(t,x)^3+\nu\frac{\partial^2 u(t,x)}{\partial x^2}
\bigg), 
\lb{eqn:PDE1}
\eeq
where $\alpha,\nu (\in\mathbb{R})$ are parameters, 
$\ve (\in\mathbb{R})$ is the small parameter,  
and $r(\in\mathbb{R})$ represents the time-delay.
Although the system is described by a delay
partial differential equation, our prescription is not changed.

First, the naive perturbation solution, 
$u(t,x)=u^{(0)}(t,x)+\ve u^{(1)}(t,x)+{\cal O}(\ve^2)$, is obtained by solving 
the following equations,
\beqa
L_r u^{(0)}(t,x)&=&0,\qquad
L_r u^{(1)}(t,x)=\alpha u^{(0)3}(t,x)+\nu\frac{\partial^2 u^{(0)}(t,x)}{\partial x^2},\non\\
L_r u(t,x)&:=&\frac{\partial u(t,x)}{\partial t}+\frac{\pi u(t-r,x)}{2r}.\non
\eeqa
The solutions are given by
\beqa
u^{(0)}(t,x)&=&A(x) \e^{i\pi t/(2r)}+\cc,\non\\
u^{(1)}(t,x)&=&\frac{3\alpha|A(x)|^2A(x)+\nu\frac{\partial^2A(x)}{\partial x^2}}{1+i\frac{\pi}{2}}t\e^{i\pi t/(2r)},\non
\eeqa
where $A(x)(\in\mathbb{C})$ is an arbitrary differentiable function of $x$.
The naive perturbation solution includes the secular term.

Next, to remove the secular behavior, 
the renormalized variable $\tA(t,x)(\in\mathbb{C})$ is defined as
\beq
\tA(t,x):=A(x)+\ve\frac{t}{1+i\frac{\pi}{2}}
\bigg(
3\alpha|A(x)|^2A(x)+\nu\frac{\partial^2 A(x)}{\partial x^2}
\bigg).
\non
\eeq

Finally, the renormalization equation, which $\tA(t,x)$ should satisfy, 
is derived as
\beqa
\frac{\partial \tA(t,x)}{\partial t}=\frac{\ve}{1+i\frac{\pi}{2}}
\bigg(
3\alpha|\tA(t,x)|^2\tA (t,x) + \nu \frac{\partial^2 \tA(t,x)}{\partial x^2}
\bigg).
\non
\eeqa
This is the well-known complex Ginzburg-Landau equation.

\subsection{Spatially extended system (2)}
To show that the renormalization method is useful 
in the case that a spatially extended system modeled by including perturbation 
terms with delayed arguments, we consider the following system,
\beq
\bigg(
\frac{\partial^2}{\partial t^2}-
\frac{\partial^2}{\partial x^2}+1
\bigg)u(t,x)=\ve a u^3(t-r,x)
\lb{eqn:PDE2},
\eeq
where $a(\in\mathbb{R})$ is a parameter, $r(\in\mathbb{R})$ is  
the time-delay, and $\ve(\in\mathbb{R},|\ve|\ll 1)$ is the small parameter.
Along with the procedure for partial differential equations proposed 
in Ref.\citen{GMN99}, we can derive the reduced equation.

First, the naive perturbation solution  
$u(t,x)=u^{(0)}(t,x)+\ve u^{(1)}(t,x)+{\cal O}(\ve^2)$ 
is obtained by solving the following equations,
\beqa
Lu^{(0)}(t,x)&=&0,\quad Lu^{(1)}(t,x)=au^{(0)3}(t-r,x),
\non\\
Lu(t,x)&:=&(\partial^2_{tt}-\partial^2_{xx}+1)u(t,x).\non
\eeqa
We set $u^{(0)}(t,x)$ as 
$$
u^{(0)}(t,x)=A\e^{i(kx-\omega t)}+\cc,
$$
where $A(\in\mathbb{C})$ is the integration constant, and 
$\omega^2=k^2+1$, with $k(\in\mathbb{R})$ being a parameter. 
A secular solution of $u^{(1)}(t,x)$ is found to be 
$$
u^{(1)}(t,x)=3a|A|A\e^{i\omega r}(p_{10}t+p_{01}x)\e^{i(kx-\omega t)}+\cc,
$$
where $p_{10}(\in\mathbb{C})$ and $p_{01}(\in\mathbb{C})$ are parameters.
The values of these parameters are restricted by the condition
\beq
2(\omega p_{10} +k p_{01}) = i.
\lb{eqn:PDE2-cond}
\eeq

Next, the renormalized variables $\tA(t,x)(\in\mathbb{C})$ is defined as 
\beq
\tA(t,x):=A+\ve 3a|A|^2A\e^{i\omega r}(p_{10}t + p_{01}x).
\lb{eqn:PDE2-RGT}
\eeq
From this definition \fr{eqn:PDE2-RGT}, we have
\beqa
\partial_t \tA(t,x)&=&3a\ve|\tA(t,x)|^2\tA(t,x)\e^{i\omega r}p_{10},
\lb{eqn:PDE2-RGt}\\
\partial_x \tA(t,x)&=&3a\ve|\tA(t,x)|^2\tA(t,x)\e^{i\omega r}p_{01}.
\lb{eqn:PDE2-RGx}
\eeqa
Finally, eliminating $p_{10}$ and $p_{01}$ 
from Eqs.\fr{eqn:PDE2-cond}, \fr{eqn:PDE2-RGt} and \fr{eqn:PDE2-RGx},
we have the following renormalization equation,
$$
\bigg(
\frac{\partial}{\partial t}+\frac{d\omega}{dk}\frac{\partial}{\partial x}
\bigg)\tA(t,x)=i\frac{3a}{2} \ve |\tA(t,x)|^2\tA(t,x)\e^{i\omega r}.
$$

\section{Extended Renormalization Method}
\lb{sec:extended}
In this section, 
we propose an extended renormalization method which can lead
to a reduced equation with 
delay from a given system with large- or order-one delay. 
In this paper, by large-delay we mean 
that the delayed arguments are of order $1/\ve^{\alpha}, (\alpha>0)$ with 
$\ve$ being the small parameter associated with the 
given weakly nonlinear system. 
We show that our reduced equations are consistent with those 
obtained by the multiple-scale method in Refs.\citen{DC05,PE00}. 

\subsection{Linear system }
The model which we study here is 
\beq
\frac{d^2 x(t)}{dt^2}+\omega^2 x(t)
+\ve x\bigg(t-\frac{r}{\ve^{\alpha}}\bigg)=0,
\lb{eqn:large-linear1}
\eeq
where $r/\ve^{\alpha}$ represents large-delay with $\ve(\in\mathbb{R})$ 
being a small parameter, and $\alpha(\geq 0)$ is a parameter.   
When $\alpha=0$, this equation is the same as Eq.\fr{eqn:linear-DDE}. 
When $\alpha=1$ and $\omega=1$, the system \fr{eqn:large-linear1} 
was analyzed using the multiple-scale method in Ref\citen{DC05}.
We analyze this system \fr{eqn:large-linear1} 
using the extended renormalization method
and compare the result with that in the previous work.

First, we obtain the perturbation solution,
$x(t)=x^{(0)}(t)+\ve x^{(0)} (t)+\ve^2 x^{(0)}(t)+{\cal O}(\ve^{3})$, 
by solving the following equations
\beqa
Lx^{(0)}(t)&=&0,
\lb{eqn:large-linear1-0}\\
Lx^{(1)}(t)&=&-x^{(0)}\bigg(t-\frac{r}{\ve^{\alpha}}\bigg),
\lb{eqn:large-linear1-1}\\
Lx^{(2)}(t)&=&-x^{(1)}\bigg(t-\frac{r}{\ve^{\alpha}}\bigg),\quad
Lx(t):=\bigg(\frac{d^2}{dt^2}+\omega^2\bigg)x(t).\non
\eeqa
In deriving these equations, the magnitude of 
$r/\ve^{\alpha}$ in the delayed argument is treated as a large value, 
this treatment corresponds to the nonstandard expansion 
in the previous study\cite{DC05}.

The solution of Eq.\fr{eqn:large-linear1-0} is 
\beq
x^{(0)}(t)=\cA(0)\exp(i\omega t)+\cc,
\lb{eqn:large-linear1-NS0}
\eeq
where $\cA(0)(\in\mathbb{C})$ represents the contribution to the solution 
$x^{(0)}(t)$ except for the fast motion $\exp (i\omega t)$.  
We assume that the solution $x^{(0)}$ at $t-r/\ve^{\alpha}$ is expressed as 
$$
x^{(0)}\bigg(t-\frac{r}{\ve^{\alpha}}\bigg)=
\cA\bigg(-\frac{r}{\ve^{\alpha}}\bigg)\e^{-i\omega r/\ve^{\alpha}}\exp(i\omega t)+\cc.
$$
Here $\cA(-r/\ve^{\alpha})$ represents  the contribution to 
$x^{(0)}(t-r/\ve^{\alpha})$, 
except for $\e^{-i\omega r/\ve^{\alpha}}\exp(i\omega t)$. 
This implies that the argument of $\cA(t)$ is only affected by 
a large time shift.
At this stage, we do not know the functional form 
of $\cA(t)$.  
The existence of $\cA(t)$ is the most fundamental assumption in this extended 
method. 
It is noted that the equation which $\cA(t)$ should perturbatively satisfy 
is our extended renormalization equation. This extended reduced equation 
is constructed by removing the secular behavior coming from the 
resonance between the frequency in the operator $L$ and 
$\propto\exp(i\omega t)$ in the forcing terms. 
When the delay $r$ becomes zero, 
the extended method corresponds to the conventional method.
Substituting this solution $x^{(0)}(t)$ into  Eq.\fr{eqn:large-linear1-1}, 
we obtain
$$
Lx^{(1)}(t)=
-\cA\bigg(-\frac{r}{\ve^{\alpha}}\bigg)\e^{-i\omega r/\ve^{\alpha}}\exp(i\omega t)+\cc.
$$
The solution is given by
$$
x^{(1)}(t)=\frac{it}{2\omega}\cA\bigg(-\frac{r}{\ve^{\alpha}}\bigg)\e^{-i\omega r/\ve^{\alpha}}
\exp(i\omega t)+\cc.
$$
At the next order in $\ve$, we obtain 
$$
x^{(2)}(t)=
\bigg\{
\frac{-t^2}{8\omega^3}+
\bigg(\frac{-i}{8\omega^3}+\frac{r}{4\omega\ve^{\alpha}}\bigg)t
\bigg\}
\cA\bigg(-\frac{2r}{\ve}\bigg)\e^{-2\omega r/\ve^{\alpha}}\exp(i\omega t)+\cc.
$$
We observe the secular behavior as we have already seen in the case that 
the delay is not large.

Next, we remove the secular behavior. To do this, we define the extended 
renormalized variable,
\beqa
\tcA(t)&:=&\cA(0)+\ve\frac{it}{2\omega}\cA\bigg(-\frac{r}{\ve^{\alpha}}\bigg)
\e^{-\omega r/\ve^{\alpha}}\non\\
&&+\ve^2 \bigg\{
\frac{-t^2}{8\omega^3}+
\bigg(\frac{-i}{8\omega^3}+\frac{r}{4\omega\ve^{\alpha}}\bigg)t
\bigg\}
\cA\bigg(-\frac{2r}{\ve^{\alpha}}\bigg)\e^{-2\omega r/\ve^{\alpha}}.
\lb{eqn:large-linear1-RGT}
\eeqa
This definition is a form of a near-identity transformation 
at the {\it function} $\cA(0)$, instead of that at the {\it constant} $A$ 
in the conventional renormalization method.
From this definition of $\tcA(t)$, we obtain 
\beqa
&&\frac{\tcA(t+\sigma)-\tcA(\sigma)}{\sigma}=\frac{\ve i}{2\omega}\cA
\bigg(
-\frac{r}{\ve^{\alpha}}\bigg)\e^{-\omega r/\ve^{\alpha}}\non\\
&&+\ve^2 
\bigg\{
\frac{-2t}{8\omega^3}+\bigg(\frac{-i}{8\omega^3}+\frac{r}{4\omega\ve^{\alpha}}\bigg)
\bigg\}
\tA\bigg(-\frac{2r}{\ve^{\alpha}}\bigg)\e^{-2\omega r/\ve^{\alpha}}
+{\cal O}(\sigma,\ve^3),
\lb{eqn:large-linear1-diff}
\eeqa
where $\sigma(\in\mathbb{R})$ is a parameter whose value is of smaller than 
$r/\ve^{\alpha}$.
The inverse of Eq.\fr{eqn:large-linear1-RGT} is derived as 
$$
\cA(0)=\tcA(t)-\ve\frac{it}{2\omega}\tcA\bigg(t-\frac{r}{\ve^{\alpha}}\bigg)
\e^{-i\omega r/\ve^{\alpha}}+{\cal O}(\ve^2).
$$
Using the above expression, we obtain the following relation 
\beqa
\cA\bigg(-m\frac{r}{\ve^{\alpha}}\bigg)&=&
\tcA\bigg(t-m\frac{r}{\ve^{\alpha}}\bigg)\non\\
&&-\ve\frac{i(t-mr/\ve^{\alpha})}{2}\tcA\bigg(t-m\frac{r}{\ve^{\alpha}}\bigg)
\e^{-i\omega r/\ve^{\alpha}}+{\cal O}(\ve^2).
\lb{eqn:large-linear1-IRGT}
\eeqa
with $m(\in\mathbb{N})$.  
We substitute Eq.\fr{eqn:large-linear1-IRGT} into 
Eq.\fr{eqn:large-linear1-diff} and take the limit $\sigma\to 0$, 
we obtain our extended renormalization method which $\tcA(t)$ should 
perturbatively satisfy,
\beq
\frac{d\tcA(t)}{dt}=\ve\frac{i}{2\omega}\e^{-i\omega r/\ve^{\alpha}}
\tcA\bigg(t-\frac{r}{\ve^{\alpha}}\bigg) 
+\ve^2\frac{-i}{8\omega^3}\e^{-2i\omega r/\ve^{\alpha}}
\tcA\bigg(t-2\frac{r}{\ve^{\alpha}}\bigg)
\lb{eqn:large-linear1-RGE}.
\eeq
In Eq.\fr{eqn:large-linear1-RGE} there are delay terms,  and 
this renormalization equation in the case of $\alpha=1$ and $\omega=1$ 
is equivalent to reduced equations derived in 
Ref.\citen{DC05}, where numerical simulation and some analysis 
have shown that the reduced system reproduces 
the behavior of slow motion in the original system.   
In the case that $\alpha=0$ and $r$ is given by Eq.\fr{eqn:linear-cond}, 
we can show that 
one of the solutions to Eq.\fr{eqn:large-linear1-RGE} up to ${\cal O}(\ve^2)$ 
is given by Eq. \fr{eqn:linear-sol-tA}.

Here we compare this extended renormalization method with the conventional 
method discussed in \S\ref{sec:conventional} for 
this system\fr{eqn:large-linear1}.
When we use the conventional method 
we cannot obtain a reduced equation. 
To see this, we use the conventional method. The naive perturbation solutions
are
\beqa
x^{(0)}(t)&=&A\e^{i\omega t}+\cc\non\\
x^{(1)}(t)&=&\frac{it}{2\omega}A\e^{-i\omega r/\ve^{\alpha}}\e^{i\omega t}+\cc\non\\
x^{(2)}(t)&=&\bigg\{ \frac{-t^2}{8\omega^3}
 +\bigg(\frac{-i}{8\omega^3}+\frac{r}{4\omega\ve^{\alpha}}\bigg)t\bigg\}
A\e^{-2i\omega r/\ve^{\alpha}}\e^{i\omega t}\non\\
&&\qquad +\cc\non
\eeqa
The renormalized variable is defined as
\beqa
&&\tA(t):=A+\ve\frac{it}{2\omega}A\e^{-i\omega r/\ve^{\alpha}}
+\ve^2
\bigg\{\frac{-t^2}{8\omega^3}+\bigg(\frac{-i}{8\omega^3}+\frac{r}{4\ve^{\alpha}}\bigg)t\bigg\}
A\e^{-2i\omega r/\ve^{\alpha}}.\non
\eeqa
The renormalization equation up to ${\cal O}(\ve)$ becomes
\beq
\frac{d\tA(t)}{dt}=\ve\frac{i}{2\omega}\e^{-i\omega r/\ve^{\alpha}}\tA(t),
\lb{eqn:large-linear1-wrongO1}
\eeq
and that up to ${\cal O}(\ve^2)$ becomes 
\beq
\frac{d\tA(t)}{dt}=\ve\frac{i}{2\omega}\e^{-i\omega r/\ve}\tA(t)+\ve^2
\bigg(\frac{r}{4\ve^{\alpha}}-\frac{i}{8\omega^3}\bigg)\e^{-2i\omega r/\ve^{\alpha}}\tA(t).
\lb{eqn:large-linear1-wrongO2}
\eeq
Since the magnitudes of the terms calculated as higher-order correction in 
Eq.\fr{eqn:large-linear1-wrongO2} 
are ${\cal O}(\ve^2)$ and ${\cal O}(\ve^{2-\alpha})$,
this approximation is in contradiction with Eq.\fr{eqn:large-linear1-wrongO1} except for 
the case of $\alpha=0$. When $\alpha=0$,
Eq.\fr{eqn:large-linear1-wrongO2} becomes Eq.\fr{eqn:linear-RGE}, and 
there is no contradiction only in the case $\alpha=0$.
We conclude that, for systems with large-delay, the extended renormalization
method should be used.

\subsection{Nonlinear system}
We consider a weakly nonlinear system with large-delay which appears in optics.
In Ref.\citen{PE00}, 
they have analyzed the system with optoelectronic feedback, 
and the system is described as
\beqa 
\frac{dx(s)}{ds}&=&-y(s)-\ve^2 x(s)
\bigg(1+\frac{2P}{1+2P}y(s)
\bigg),
+\ve^2 C\bigg\{1+y\bigg(s-\frac{\Theta}{\ve^2}\bigg)\bigg\}\non\\
\frac{dy(s)}{ds}&=&(1+y(s))x(s),\non
\eeqa
where $s$ is the scaled time $C,P,\Theta(\in\mathbb{R})$ are parameters, 
and $\ve(\in\mathbb{R})$ is the small parameter.
The solution which we focus on is the small amplitude regime, described by the 
following assumption
\beqa
x(s)&=&\ve x^{(1)}(s)+\ve^2 x^{(2)}(s)+\ve^3 x^{(3)}(s)+{\cal O}(\ve^4),\non\\
y(s)&=&\ve y^{(1)}(s)+\ve^2 y^{(2)}(s)+\ve^3 y^{(3)}(s)+{\cal O}(\ve^4).\non
\eeqa
We construct the reduced equation using our extended method, and compare the
result with that reported in Ref.\citen{PE00}.
 
First the naive perturbation problems are
\beqa 
\frac{dx^{(1)}(s)}{ds}&=&-y^{(1)}(s),\quad  \frac{dy^{(1)}(s)}{ds}=x^{(1)}(s),\non\\
\frac{dx^{(2)}(s)}{ds}&=&-y^{(2)}(s)+C,\quad
\frac{dy^{(2)}(s)}{ds}=x^{(2)}(s)+x^{(1)}(s)y^{(1)}(s),\non\\
\frac{dx^{(3)}(s)}{ds}&=&-y^{(3)}(s)-x^{(1)}(s)+Cy^{(1)}(s-\theta),\non\\
\frac{dy^{(3)}(s)}{ds}&=&x^{(3)}(s)+y^{(2)}(s)x^{(1)}(s)+y^{(1)}(s)x^{(2)}(s),
\non
\eeqa
where $\theta=\Theta/\ve^2$.
The solutions $x^{(1)}(s),x^{(2)}(s)$ and $x^{(3)}(s)$ are given by 
\beqa
x^{(1)}(s)&=&\cA(0)\e^{is}+\cc,
\quad x^{(2)}(s)=\frac{-i}{3}\cA(0)^2\e^{2is}+\cc,
\non\\
x^{(3)}(s)&=&\frac{s}{2}\bigg( iC\cA(0)-\frac{i}{3}|\cA(0)|^2\cA(0)-\cA(0)
\non\\
&&-iC\cA(-\theta)\bigg)\e^{is}+\cc+\mbox{higher harmonics}.\non
\eeqa

The definition of the renormalized variable is 
\beqa
\tcA(t)&:=&\cA(0)+\ve^2\frac{s}{2}
\bigg( iC\cA(0)-\frac{i}{3}|\cA(0)|^2\cA(0)-\cA(0)
-iC\cA(-\theta)\bigg).
\lb{eqn:large-nonlinear1-RGD}
\eeqa
The renormalization equation is derived from Eq.\fr{eqn:large-nonlinear1-RGD} 
as
\beq
\frac{d\tcA(s)}{ds}=\frac{\ve^2}{2} 
\bigg( iC\tcA(s)-\frac{i}{3}|\tcA(s)|^2\tcA(s)-\tcA(s)
-iC\tcA(s-\theta)\bigg).
\lb{eqn:large-nonlinear1-RGE}
\eeq
The renormalization equation \fr{eqn:large-nonlinear1-RGE} is the 
reduced equation derived in Ref.\citen{PE00}.
Some analytical analyzes in Ref.\citen{PE00} have shown where 
bifurcation points are. Again, we confirm that our extended method gives
the same results given by the use of the multiple-scale method.

\section{Discussion and Conclusions}
\lb{sec:conclusion}
In this section, we discuss the features of both our conventional and 
extended renormalization methods, and then we conclude our study.

The prescription of the conventional renormalization method 
for systems with order-one delay is not different from ones without delay. 
The standard prescription leads to the reduced equation from a given 
weakly nonlinear system.   
The conventional method removes the secular terms from naive perturbation 
series by accounting for their effect with renormalized variables.
Derived reduced systems are always ones without delay.
Being without delay in a reduced equation means that 
the dimension of phase space for 
the original system can be reduced perturbatively. This reduction provides 
us the approximate structure of phase space.
Systems to which we can apply this method 
are weakly nonlinear ones with order-one delay. In this sense, 
the conventional method is restricted.

The prescription of our proposed extended renormalization method also removes 
the secularity. The basic assumption for this method is that 
we can introduce an unknown function contributing the naive perturbation 
solution, instead of the integration constant 
in the use of the conventional method.
Although a rigorous mathematical meaning of the extended method 
has not yet given in this paper, 
we have checked the validity of our method through various examples.  
Derived reduced systems using this extended method are 
always ones with delay. 
This means that the dimensions of phase space for both the original and 
reduce systems are high. 
The advantage of our reduction method is that 
a steady state in the reduced system corresponds to a 
periodic one in the given system, which provides us some bifurcation analyzes.
Using extended method, we can deal with 
systems whose delay time is of order $1/\ve^{\alpha}$, $(\alpha\geq 0)$ 
where $\ve$ is the 
small parameter appearing in the original system under study.

For both the renormalization methods, 
terms in the reduced equation arise from 
secular terms appearing in the naive perturbation analysis. 
This implies that higher harmonics in the 
naive perturbation analysis does not contribute to the reduced equation 
in the first order approximation. 
In this sense, the reduced equation can be obtained from 
a wide class including the original system. 
Compared to the multiple-scale method, our methods do not need scaled 
variables.
While in the course of the derivation of a reduced 
system using our methods, we need the analytical expressions of the naive  
perturbation solutions so that we define the renormalized variables.
Although the procedures of our methods are systematic, 
the application of our methods are restricted by this disadvantage.

In this paper, we have shown that the renormalization method can be extended 
to a tool to study systems with delay, 
and that the method gives reduced systems successfully. 
Combining the previous studies of 
the renormalization method
with the present study, we expect that  
our renormalization method includes all the asymptotic analyzes.  
Furthermore, 
we believe that the application of the renormalization method can help 
elucidate the behavior of time-delayed systems in a non-chaotic regime.

\section*{Acknowledgements}
The author would like to thank the members of 
NTT Communication Science Laboratories for 
their continual encouragement.

%

\end{document}